# Ring Optimized M-APSK Modulation for Discrete Modulated CV-QKD

Seonguk Kim, Jun Heo*
School of Electrical Engineering, Korea University, Seoul, Republic of Korea.

*Corresponding author. E-mail(s): junheo@korea.ac.kr;
Contributing authors: djm06145@korea.ac.kr;

## Abstract

This paper proposes a multi ring M-APSK constellation optimization method for discrete-modulated continuous variable quantum key distribution. Unlike conventional APSK structures with fixed ring spacing and predefined ring probabilities, the proposed method optimizes the ring radius ratio and ring probability to improve the finite-size secret key rate. A method based on the Gram matrix is used to calculate the nonzero spectrum of the average state $\tau$, and fidelity is used to compare the optimized discrete average state with the Gaussian average state. The results show that the proposed structure extends the maximum transmission distance of 16-APSK by approximately 15% compared with the conventional binomial APSK structure. The optimization gain is larger for small size APSK constellations, where the average state has a larger structural gap from Gaussian modulation.

**Keywords:** Quantum key distribution, Continuous variable QKD, Non-Gaussian QKD, constellation

# 1 Introduction

Quantum key distribution (QKD) allows Alice and Bob to share a secret key through a quantum channel and an authenticated classical channel. Unlike conventional cryptographic systems based on computational hardness, QKD relies on the principles of quantum mechanics and can detect the presence of eavesdropping. QKD protocols are usually divided into discrete variable QKD (DV-QKD) and continuous variable QKD (CV-QKD), depending on the physical variables used to encode quantum information [1, 2].

CV-QKD uses the quadrature of light, the amplitude and phase components, to transmit information. Since it can be implemented with coherent states and coherent detection, CV-QKD is highly compatible with existing optical components and coherent optical communication systems. This makes CV-QKD a promising structure for practical optical network applications [2-5]. Gaussian modulation is the standard modulation format in CV-QKD. It provides a high theoretical secret key rate, and the security analysis of Gaussian states can be efficiently performed using covariance matrices [6-8]. However, ideal Gaussian modulation is difficult to implement exactly because practical optical modulators have finite resolution and dynamic range, while continuously distributed amplitudes must be generated with high precision [9]. Discrete-modulated CV-QKD addresses this limitation by restricting Alice's signals to a finite set of coherent states, which can simplify state preparation and practical signal processing [10-12]. Security analyses have subsequently been developed for low-order and more general finite constellations [13-16]. Under comparable channel conditions, however, finite-constellation modulation generally achieves a lower secret key rate than Gaussian modulation. Therefore, it is important to design finite constellations that approximate the relevant properties of Gaussian modulation as closely as possible [15-17]. Early studies on discrete-modulated CV-QKD mainly focused on including 2-PSK, 4-PSK, and 8-PSK [10-14]. Because single-ring PSK provides limited freedom in the amplitude dimension, multi-amplitude formats such as QAM and APSK have been investigated to provide greater flexibility in constellation design [17-19].

Recent studies have investigated constellation shaping as a means of improving the performance of DM-CV-QKD with finite coherent-state constellations. Probabilistic shaping has been applied to QAM and APSK constellations by assigning nonuniform selection probabilities to the transmitted coherent states [18–20]. Geometric shaping and multi-amplitude constellation design have also been explored through amplitude-phase and multi-ring constellations [21, 22]. Experimental and implementation-oriented studies have further investigated probabilistically shaped APSK transmission, shaped-constellation validation, high-performance long-distance DM-CV-QKD, and transmitter-side constellation pre-optimization [18, 22–24]. Nevertheless, existing studies have primarily considered predefined probability distributions or specific two-ring, QAM, and high-order APSK formats. The joint optimization of ring geometry and ring probability across multiple multi-ring APSK orders, together with a quantitative analysis of how the average-state structure is related to the resulting key-rate improvement, remains insufficiently explored.

In this work, we extend the conventional multi-ring M-APSK framework by jointly optimizing the relative ring ratios, ring probabilities, and modulation variance under the same secret-key rate model used for the conventional baseline [17]. The optimization is applied to 16, 32, and 64-APSK constellations to investigate how the achievable shaping gain varies with constellation size. Changes in the constellation geometry and probability distribution modify the average modulation state $\tau$, which in turn affects the correlation parameter $Z^*$ entering the secret key rate calculation [15]. To evaluate the spectral quantities of $\tau$required for computing $Z^*$, we construct its nonzero spectral decomposition using an $M \times M$ Gram-matrix representation. In addition, the fidelity between the discrete average state and the Gaussian thermal state is employed as a complementary structural metric to quantify how the optimization changes the similarity of the average state to Gaussian modulation.

# 2 Secret key rate

## 2.1 Finite-size secret key rate

In this paper, we use the finite-size secret key rate to compare the performance of different M-APSK constellation structures. In a practical CV-QKD system, Alice and Bob exchange only a finite number of quantum states. Therefore, part of the transmitted data is used for channel parameter estimation, and the remaining data is used for information reconciliation. The finite-size secret key rate is written as [17, 25]:

$$K = \frac{m}{N}\left(\beta I_{BA}(V_A, T_{min}, \xi_{max}) - \chi_{BE}(V_A, T_{min}, \xi_{max}, Z^*) - \Delta m\right) \tag{1}$$

where $N$ is the total number of transmitted states and $m$ is the number of states used for information reconciliation. $\beta$ is the reconciliation efficiency, and $I_{BA}$ is the mutual information between Bob and Alice. $\chi_{BE}$ denotes the Holevo information that Eve can obtain about Bob's key, and $\Delta m$ is the correction term caused by privacy amplification in the finite-size regime. In the finite-size regime, the channel transmittance $T$ and the excess noise $\xi$ cannot be known exactly. Therefore, conservative values obtained from parameter estimation are used. In this work, the lower bound of the channel transmittance, $T_{min}$, and the upper bound of the excess noise, $\xi_{max}$, are used in the secret key rate calculation. In other words, $T$ and $\xi$ are replaced by $T_{min}$ and $\xi_{max}$ in practical calculation. The mutual information $I_{BA}$ is determined by the modulation variance $V_A$, the channel transmittance, and the excess noise. On the other hand, the Holevo information $\chi_{BE}$ also depends on the correlation parameter $Z^*$, which represents the correlation between Alice and Bob. Therefore, when the constellation structure changes, the average state $\tau$ changes, and this changes $Z^*$ and the final secret key rate.

In this paper, we use the same finite-size secret key rate model and simulation parameters as the previous multi-ring M-APSK study [17]. Therefore, the purpose of this work is not to modify the secret key rate model itself, but to analyze how the secret key rate changes when the ring ratio and ring probability of the M-APSK constellation are optimized under the same finite-size conditions. Recent studies have further developed the security analysis of DM-CV-QKD by addressing dimension reduction, nonideal detection, composable finite-size security, and coherent attacks [26-29]. In this work, however, we use the same finite-size secret key rate model and simulation parameters as the previous multi-ring M-APSK study [17] to isolate the performance change caused by constellation optimization.

## 2.2 Constellation average state and Parameter Z

An important part of the secret key rate calculation is the parameter $Z^*$, which represents the correlation between Alice and Bob. Unlike Gaussian modulation, in discrete modulation it is difficult to directly determine the correlation parameter of the EB (entanglement based) protocol covariance matrix only from the measurement data obtained in the PM (prepare and measure) protocol. Several approaches have been developed to estimate or bound the relevant correlation term for discrete modulation [13-16]. In this paper, we use $Z^*$ from [15], which provides an explicit lower bound for arbitrary modulation.

For the Gaussian channel model considered in this work, $Z^*$is evaluated as follows.

$$Z^*(T,\xi,\tau) = 2\sqrt{T}tr\left(\tau^{\frac{1}{2}}\hat{a}\tau^{\frac{1}{2}}\hat{a}^\dagger\right) - \sqrt{2T\xi w} \tag{2}$$

$T$ is the channel transmittance, $\xi$ is the excess noise, and $\hat{a}$ and $\hat{a}^\dagger$ are the annihilation and creation operators, respectively. $w$ and $\hat{a}_\tau$ are defined as

$$w = \sum_k p_k(\langle\alpha_k|\hat{a}_\tau^\dagger\hat{a}_\tau|\alpha_k\rangle - |\langle\alpha_k|\hat{a}_\tau|\alpha_k\rangle|^2) \tag{3}$$

$$\hat{a}_\tau \ = \tau^{\frac{1}{2}}\hat{a}\tau^{-\frac{1}{2}} \tag{4}$$

As shown in the above equation, $Z^*$ depends not only on the channel parameters but also directly on the average state $\tau$. In discrete modulation, $\tau$ is defined as the density matrix of the coherent-state ensemble sent by Alice.

$$\tau = \sum_k p_k\ |\alpha_k\rangle\langle\alpha_k| \tag{5}$$

Here, $p_k$ is the probability that the k-th constellation point is selected, and $\alpha_k$ is the complex amplitude of the corresponding coherent state. The coherent state $|\alpha_k\rangle$ can be written in the Fock basis as

$$|\alpha_k\rangle = e^{-\frac{|\alpha_k|^2}{2}}\sum_{n=0}^{\infty}\frac{\alpha_k^n}{\sqrt{n!}}|n\rangle \tag{6}$$

Also, $\alpha_k$ is complex amplitude and can be expressed using its radius and phase as

$$\alpha_k = r_k e^{i\theta_k} \tag{7}$$

$$|\alpha_k\rangle = |r_k e^{i\theta_k}\rangle = e^{-\frac{r_k^2}{2}}\sum_{n=0}^{\infty}\frac{r_k^n e^{i\theta_k n}}{\sqrt{n!}}|n\rangle \tag{8}$$

Therefore, $\tau$ contains the radius $r_k$, phase $\theta_k$, and probability $p_k$ of each constellation point. When the ring ratio or ring probability is adjusted, the set $\{\alpha_k, p_k\}$ changes, and this modifies τ. This modification is reflected in the final secret key rate through $Z^*$. The relation considered in this paper can be summarized as $\{\alpha_k, p_k\} \rightarrow \tau \rightarrow Z^* \rightarrow$ Secret key rate.

## 2.3 Gram Matrix Method

To compute $Z^*$, the eigen structure of the average state $\tau$ is required because $\tau^{\frac{1}{2}}$ and $\tau^{-\frac{1}{2}}$ appear in the expression of $Z^*$. For one-ring PSK, a closed-form decomposition can be obtained by exploiting its rotational symmetry [11, 15]. However, for multi-ring APSK, a general closed form expression is difficult to obtain. The average state of multi-ring APSK can be written as the sum of the average states of each ring,

$$\tau_D = \tau_1 + \tau_2 + \cdots + \tau_R \tag{9}$$

Even though each ring state $\tau_R$ has its own eigen structure, the eigen bases of different rings are generally not the same. Therefore, the eigen structure of the total state $\tau_D$ cannot be directly obtained from the eigenstructures of the individual ring states. In the conventional approach, $\tau_D$ is diagonalized directly in the Fock basis.

Since a coherent state is represented by an infinite sum in the Fock basis, a sufficiently large photon-number cut-off is used in numerical calculations to approximate $\tau_D$. Let $N_F$ denote the dimension of the truncated Fock-space representation. Then, the conventional approach diagonalizes $\tau_D$ as an $N_F \times N_F$ matrix. However, when the discrete constellation consists of $M$ coherent states, the number of non-zero eigenvalues of $\tau_D$ is at most $M$. Thus, direct diagonalization in the Fock basis computes the eigen structure in a space larger than the effective support of $\tau_D$.

In this paper, we use the Gram matrix method to handle this effective support directly. The Gram matrix is defined by the inner products between constellation points, and it has the same nonzero eigenvalues as $\tau_D$. Therefore, the nonzero spectrum of $\tau_D$ can be obtained from an $M \times M$ Gram matrix, instead of diagonalizing the $N_F \times N_F$ Fock-space matrix. The Gram matrix method is as follows. For a discrete constellation composed of $M$ coherent states, we define the matrix $V$ by collecting the weighted coherent states:

$$V = \left[\sqrt{p_1}|\alpha_1\rangle, \sqrt{p_2}|\alpha_2\rangle, \dots \sqrt{p_M}|\alpha_M\rangle\right] \tag{10}$$

Then, the average state $\tau$ can be written as $\tau = VV^{\dagger}$. The corresponding Gram matrix $G$ is defined as $G = V^{\dagger}V$. Each element of the Gram matrix is given by the overlap between coherent states and their probabilities:

$$G_{ij} = \langle v_i | v_j \rangle = \sqrt{p_i}\sqrt{p_j}\langle \alpha_i | \alpha_j \rangle \tag{11}$$

The important property is that $VV^{\dagger}$ and $V^{\dagger}V$ have the same non-zero eigenvalues. Therefore, the non-zero eigenvalues of $\tau$ can be obtained from the $M \times M$ Gram matrix $G$. If the eigenvalue and eigenvector of $G$ are denoted by $\lambda_j$ and $u_j$, respectively, then

$$V^{\dagger}Vu_j = \lambda_j u_j \tag{12}$$

$$VV^{\dagger}(Vu_j) = \lambda_j(Vu_j) \tag{13}$$

Thus, $\lambda_j$ is also a nonzero eigenvalue of $\tau$, and $Vu_j$ gives the corresponding eigenvector direction. The corresponding eigenvector of $\tau$ is obtained as

$$|\phi_j\rangle = \frac{1}{\sqrt{\lambda_j}} Vu_j \tag{14}$$

The average state $\tau$ can be written in the spectral decomposition form for its non-zero eigenvalues:

$$\tau = \sum_{j=0}^{M-1} \lambda_j \, |\phi_j\rangle\langle\phi_j| \tag{15}$$

Using this expression, $\tau^{1/2}$ is calculated as

$$\tau^{1/2} = \sum_{j=0}^{M-1} \sqrt{\lambda_j} \, |\phi_j\rangle\langle\phi_j| \tag{16}$$

When calculating $\tau^{-1/2}$ the inverse is taken only for the non-zero eigenvalues of $\tau$. Zero eigenvalues are not inverted. Therefore, the Gram matrix method allows us to obtain the non-zero spectrum of $\tau$ from the $M \times M$ Gram matrix, without directly diagonalizing the full truncated Fock-space representation. The Gram matrix representation reduces the dimension of the eigen decomposition, while the truncated Fock representation is retained for evaluating the required operator matrix elements.

# 3 Constellation Optimization

## 3.1 Ring Probability and Radius Adjustment

In this paper, we consider a constellation optimization method based on the conventional multi-ring M-APSK structure by adjusting the ring radius and ring probability. In the conventional M-APSK structure, the ring radius is equally spaced, and the ring probability is given by either a uniform or binomial distribution, as defined in Ref. [17]. Previous work compared the uniform and binomial distributions and showed that the binomial distribution provides a higher secret key rate. However, it is not clear whether such fixed ring spacing and fixed probability distribution are always optimal for a given CV-QKD condition. Therefore, in this paper, we keep the basic M-APSK structure and adjust only the ring spacing and ring probability [17].

First, we mainly analyze 16-APSK (4+12), which is the smallest multi-ring structure. Since a small constellation has a larger gap from Gaussian modulation, there is more room for performance improvement by adjusting the ring spacing and probability. Then, the same method is extended to 32-APSK and 64-APSK to verify the applicability of the proposed method to larger multi-ring structures. For an $R$-ring APSK constellation, the outer ring is used as the reference. The radius ratio vector and ring probability vector are defined as:

$$r = (r_1, r_2, \dots, r_{R-1}, 1), \qquad p = \left(p_1, p_2, \dots, p_{R-1}, 1 - \sum_{i=1}^{R-1} p_i\right) \tag{17}$$

Here, $r$ determines the relative spacing between rings, while the overall constellation scale is determined by $V_A$. Meanwhile, in CV-QKD, the secret key rate strongly depends on the modulation variance $V_A$. The modulation variance $V_A$ is defined as:

$$V_A = 2\langle n \rangle \tag{18}$$

$$\langle n \rangle = \sum_{k=0}^{M-1} p_k |\alpha_k|^2 \tag{19}$$

In general, when calculating the distance-dependent secret key rate, $V_A$ is optimized for each transmission distance. In this paper, we add the ring ratio $r$ and the ring probability $p$ to this optimization process. Therefore, for a given transmission distance $L$, we jointly search $V_A$, $r$ and $p$ to find the constellation that maximizes the finite-size secret key rate.

$$K_{opt}(L) = \max K_L\,(V_A, r, p) \tag{20}$$

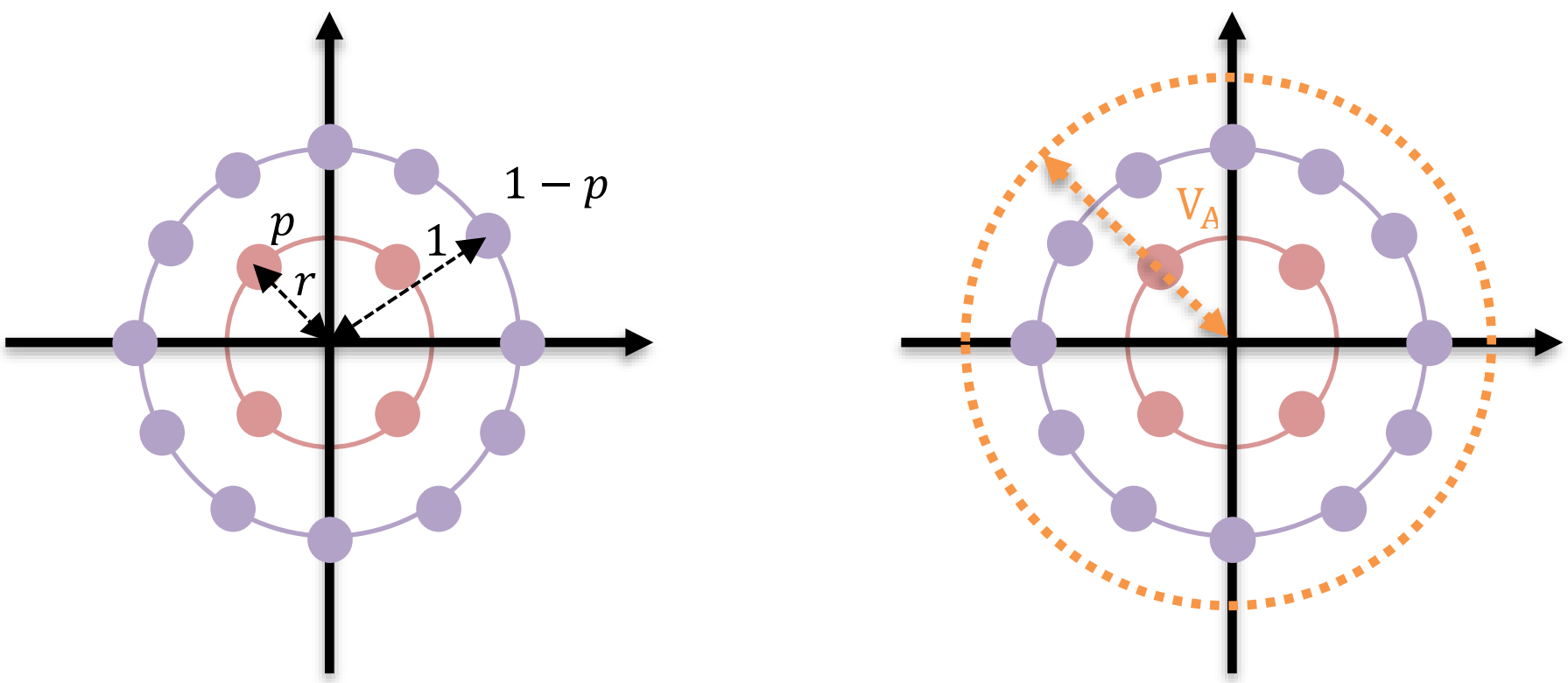

**Fig. 1** Definition of radius, probability and modulation variance for the 16 APSK (4+12) constellation.

### 3.2 Fidelity analysis of average states

In previous studies on discrete-modulated CV-QKD, the performance change caused by the constellation structure is mainly analyzed through the correlation parameter $Z^*$. The parameter $Z^*$ is directly used in the secret key rate calculation and represents the correlation between Alice and Bob. However, $Z^*$ is not determined only by the average state $\tau$ of the constellation. It also includes the channel transmittance $T$, excess noise $\xi$, and the $w$ term. Therefore, it is difficult to separate the structural effect of the constellation only from $Z^*$.

In this paper, we focus on the average state $\tau$ to compare the constellation structure itself. Here, $\tau$ is the density matrix that represents the average state defined by the modulation. For discrete modulation, $\tau_D$ is given by a mixture of a finite number of coherent states,

$$\tau_D = \sum_k p_k \, |\alpha_k\rangle\langle\alpha_k| \tag{21}$$

Here, $\alpha_k$ is the complex amplitude of the $k$-th constellation point, and $p_k$ is the probability of choosing the corresponding coherent state. Thus, $\tau_D$ contains both the positions of the constellation points and their probabilities. When the ring radius and ring probability are changed, $\tau_D$ also changes, and this change is reflected in the secret key rate through $Z^*$.

In CV-QKD, Gaussian modulation is used as an ideal reference modulation. If the average photon number is denoted by $\langle n\rangle$, the average state of Gaussian modulation is written as a thermal-state density matrix in the Fock basis [6, 15]:

$$\tau_G = \frac{1}{1+\langle n\rangle} \sum_{n=0}^{\infty} \left(\frac{\langle n\rangle}{1+\langle n\rangle}\right)^n |n\rangle\langle n| \tag{22}$$

As shown in this expression, $\tau_G$ has only diagonal terms in the Fock basis. In contrast, the density matrix $\tau_D$ of discrete modulation is a mixture of a finite number of coherent states, and it contains both diagonal and off-diagonal terms. Using the Fock-basis expression of a coherent state, the matrix element of $\tau_D$ is given by

$$\langle n|\tau_D|m\rangle = \sum_k p_k \, e^{-|\alpha_k|^2} \frac{\alpha_k^n (\alpha_k^*)^m}{\sqrt{n!\, m!}} \tag{23}$$

In particular, the diagonal term is written as

$$\langle n|\tau_D|n\rangle = \sum_k p_k\, e^{-|\alpha_k|^2} \frac{|\alpha_k|^{2n}}{n!} \tag{24}$$

This term can be interpreted as a weighted sum of Poisson-like photon-number distributions from the coherent states. Therefore, adjusting the ring radius changes the center of the photon-number distribution generated by each ring. Adjusting the ring probability changes the contribution weight of each ring.

The off-diagonal terms are related to the phase structure of the constellation. When the symbols in a ring are uniformly spaced in phase, some off-diagonal terms are canceled by phase summation. Thus, increasing the number of symbols in a ring tends to reduce the off-diagonal components. Also, increasing the number of rings adds more Poisson-like distributions with different radius, so the diagonal structure of $\tau_D$ can become closer to that of $\tau_G$.

$$\tau_G = \begin{pmatrix} \bullet & 0 & 0 & 0 & 0 & 0 & \cdots \\ 0 & \bullet & 0 & 0 & 0 & 0 & \cdots \\ 0 & 0 & \bullet & 0 & 0 & 0 & \cdots \\ 0 & 0 & 0 & \bullet & 0 & 0 & \cdots \\ 0 & 0 & 0 & 0 & \bullet & 0 & \cdots \\ 0 & 0 & 0 & 0 & 0 & \bullet & \cdots \\ \vdots & \vdots & \vdots & \vdots & \vdots & \vdots & \ddots \end{pmatrix} \quad \tau_D = \begin{pmatrix} \bullet & 0 & 0 & 0 & \bullet & 0 & \cdots \\ 0 & \bullet & 0 & 0 & 0 & \cdot & \cdots \\ 0 & 0 & \bullet & 0 & 0 & 0 & \cdots \\ 0 & 0 & 0 & \cdot & 0 & 0 & \cdots \\ \bullet & 0 & 0 & 0 & \cdot & 0 & \cdots \\ 0 & \cdot & 0 & 0 & 0 & \cdot & \cdots \\ \vdots & \vdots & \vdots & \vdots & \vdots & \vdots & \ddots \end{pmatrix}$$

**Fig. 2** Matrix structure of $\tau_G$ and $\tau_D$ in the Fock basis

Fig. 2 compares the average density matrices of Gaussian modulation and discrete modulation in the Fock basis. The Gaussian average density matrix $\tau_G$ has only diagonal terms, which follow a thermal photon-number distribution. In contrast, the discrete average density matrix $\tau_D$ is illustrated using a 4-PSK example and contains both diagonal and off-diagonal terms. The diagonal terms of $\tau_D$ are weighted sums of Poisson-like distributions from coherent states, while the off-diagonal terms are determined by the phase structure and the number of symbols in the ring. Therefore, the proposed ring radius and ring probability optimization can be interpreted as a structural adjustment that makes the discrete average state $\tau_D$ closer to the Gaussian average state $\tau_G$. To quantify this relation, we use the quantum fidelity defined as [30]:

$$F(\tau_D, \tau_G) = \left(\mathrm{Tr}\sqrt{\sqrt{\tau_G}\,\tau_D\sqrt{\tau_G}}\right)^2 \tag{25}$$

Fidelity measures the similarity between two density matrices. If $F = 1$, the two states are identical. If $F = 0$, their supports do not overlap. In this paper, we compare $\tau_D$ before and after optimization with the reference state $\tau_G$ under the same modulation variance $V_A$. This allows us to check whether the ring radius and ring probability adjustment makes the discrete density matrix closer to the Gaussian density matrix. We also analyze whether this structural change shows a consistent trend with the changes in $Z^*$ and the finite-size secret key rate.

# 4 Results

## 4.1 Two-Ring APSK

To compare the performance gain caused only by the constellation optimization, the channel and detector parameters used for the finite-size secret key rate calculation are set to be the same as those in the conventional multi-ring $M$-APSK study [17]. In this section, we compare the uniform, binomial, $r$-optimized, $p$-optimized, and $r, p$-optimized structures for 16-APSK (4+12). For each transmission distance, the modulation variance $V_A$ is optimized together with the constellation parameters. In the proposed method, the ring ratio $r$and the ring probability $p$are included as additional optimization variables. Specifically, a grid search is performed over $L = 0{:}\,0.01{:}\,60\,\text{km}$, $V_A = 0{:}\,0.01{:}\,8.0$, $r = 0{:}\,0.01{:}\,1$, and $p = 0{:}\,0.01{:}\,1$. For each distance $L$, the parameter set $(V_A, r, p)$ that maximizes the finite-size secret key rate is selected.

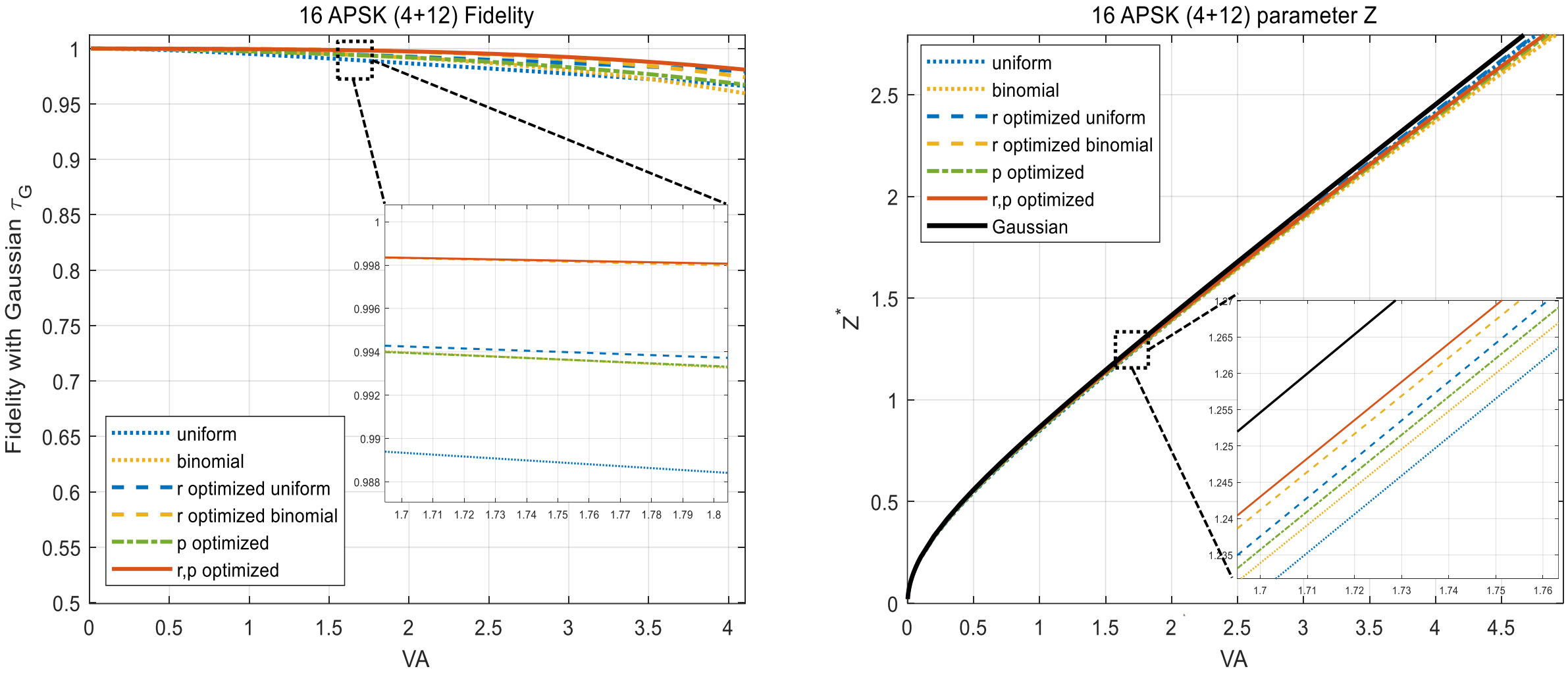

**Fig. 3** Structural analysis of 16-APSK (4+12) for modulation variance optimization at $L = 30$ km.
(a) Fidelity comparison for the conventional and optimized constellation structures.
(b) $Z^*$ comparison for the conventional, optimized, and Gaussian modulation structures.

Fig. 3(a) shows the fidelity $F(\tau_D, \tau_G)$ as a function of the modulation variance $V_A$ at a transmission distance of 30 km. The $r$-optimized and $r, p$-optimized structures show higher fidelity than the non-optimized structures. This indicates that adjusting the ring radius and ring probability makes the discrete average state $\tau_D$ closer to the Gaussian average state $\tau_G$. In addition, the binomial structure and the $p$-optimized structure show similar fidelity values. This result suggests that the conventional binomial probability is already close to a good probability shaping for 16-APSK (4+12).

Fig. 3(b) shows the correlation parameter $Z^*$ as a function of the modulation variance $V_A$ under the same conditions. The $r, p$-optimized structure shows a higher $Z^*$ than the conventional uniform and binomial structures, and its value is closer to that of Gaussian modulation. This result shows that the increase in the structural similarity of the average state, observed in Fig. 3, is consistent with the increase in $Z^*$. Fidelity is a measure of the structural similarity between $\tau_D$ and $\tau_G$, whereas $Z^*$ is the correlation parameter directly used in the secret key rate calculation. Therefore, Fig. 3 shows how the structural change of the average state caused by the optimization is reflected in $Z^*$, which is the key parameter in the secret key rate calculation.

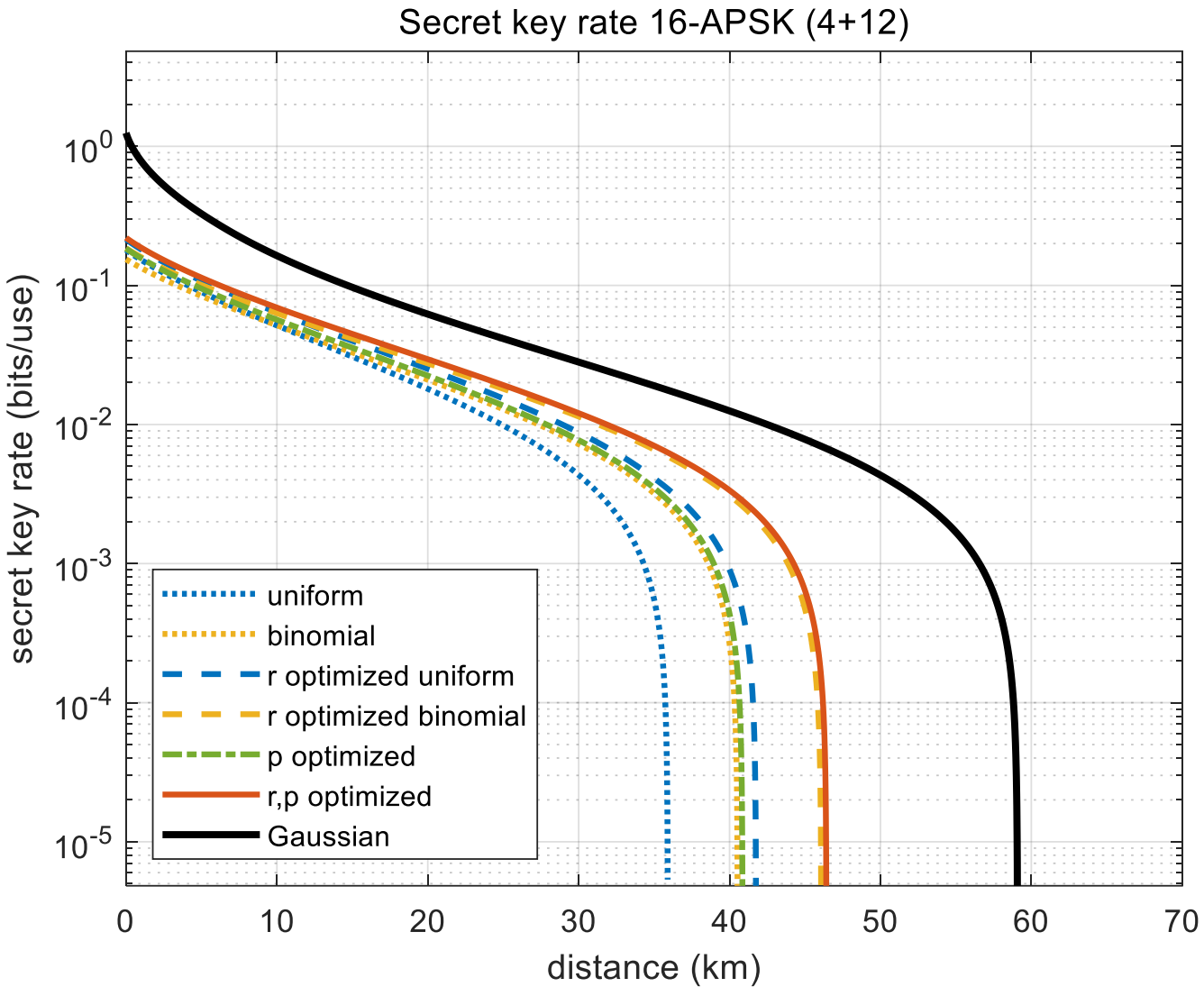

**Fig. 4** shows the finite-size secret key rate as a function of transmission distance for 16-APSK (4+12). For all structures, the modulation variance $V_A$ is optimized at each distance. In the proposed structures, the ring ratio $r$ and the ring probability $p$ are also optimized. As a result, the $r, p$-optimized structure maintains a positive secret key rate over a longer distance than the conventional uniform and binomial structures.

Fig. 4 shows, at the threshold $K = 1 \times 10^{-5}$ *bits/use*, the conventional binomial 16-APSK reaches 40.52 km, while the $r, p$-optimized structure reaches 46.39 km. This corresponds to a 14.49% improvement in transmission distance. In the considered parameter range, fidelity more clearly separates the constellation groups according to their average-state similarity, whereas $Z^*$ directly quantifies the correlation entering the security-rate calculation. Fidelity is used only as a structural diagnostic and does not replace the security analysis based on $Z^*$. This can be understood from the fact that fidelity directly compares the structural similarity between $\tau_D$ and $\tau_G$. For 16-APSK (4+12), the binomial structure shows a performance very similar to the $p$-optimized structure. However, this trend is not generally maintained in larger multi-ring structures. Therefore, the binomial probability should be interpreted as a structural feature of 16-APSK (4+12), rather than a general optimal distribution for all APSK structures.

## 4.2 Multi-Ring APSK

The ring radius and ring probability search method was applied to multi-ring APSK structures. We analyzed the 4-APSK, 16-APSK (4+12), 32-APSK (4+12+16), and 64-APSK (4+12+16+32) structures used in the conventional multi-ring study. Since 4-APSK is a one-ring structure, ring spacing and ring probability adjustment cannot be applied to it. Therefore, it is used only as a reference case. The grid search was performed over $L = 0{:}0.1{:}60$ km, $V_A = 0{:}0.1{:}8.0$, $r = 0{:}0.01{:}1$, and $p = 0{:}0.01{:}1$. A coarser grid was used for the multi-ring analysis to limit the computational cost, resulting in a slight difference from the finer-grid 16-APSK result in Section 4.1.

**Table. 1** Before optimization / After optimization

| Modulation | Case | Radius ratio | Ring probability | $V_A$ |
|---|---|---|---|---|
| 16-APSK | Conventional | [0.50, 1] | [0.75, 0.25] | 1.20 |
| 16-APSK | Proposed | [0.41, 1] | [0.71, 0.29] | 1.30 |
| 32-APSK | Conventional | [0.3333, 0.6667, 1] | [0.625, 0.3125, 0.0625] | 1.60 |
| 32-APSK | Proposed | [0.25, 0.60, 1] | [0.50, 0.43, 0.07] | 2.00 |
| 64-APSK | Conventional | [0.25, 0.50, 0.75, 1] | [0.5469, 0.3281, 0.1094, 0.0156] | 1.90 |
| 64-APSK | Proposed | [0.21, 0.43, 0.67, 1] | [0.38, 0.33, 0.25, 0.04] | 2.50 |

Table 1 summarizes the main constellation parameters of the conventional binomial structure and the $r, p$-optimized structure near the maximum transmission distance at $K = 1 \times 10^{-5}$ *bits/use*. The ring radius ratio, ring probability, and modulation variance $V_A$ are listed before and after optimization. A common trend is observed after optimization. The ring radius ratios are reduced from the equal-spacing values, which indicates that the inner rings are shifted inward. The ring probabilities also become more evenly distributed over the rings. This change is especially clear in the outermost ring, where the probability differs significantly from that of the conventional binomial distribution.

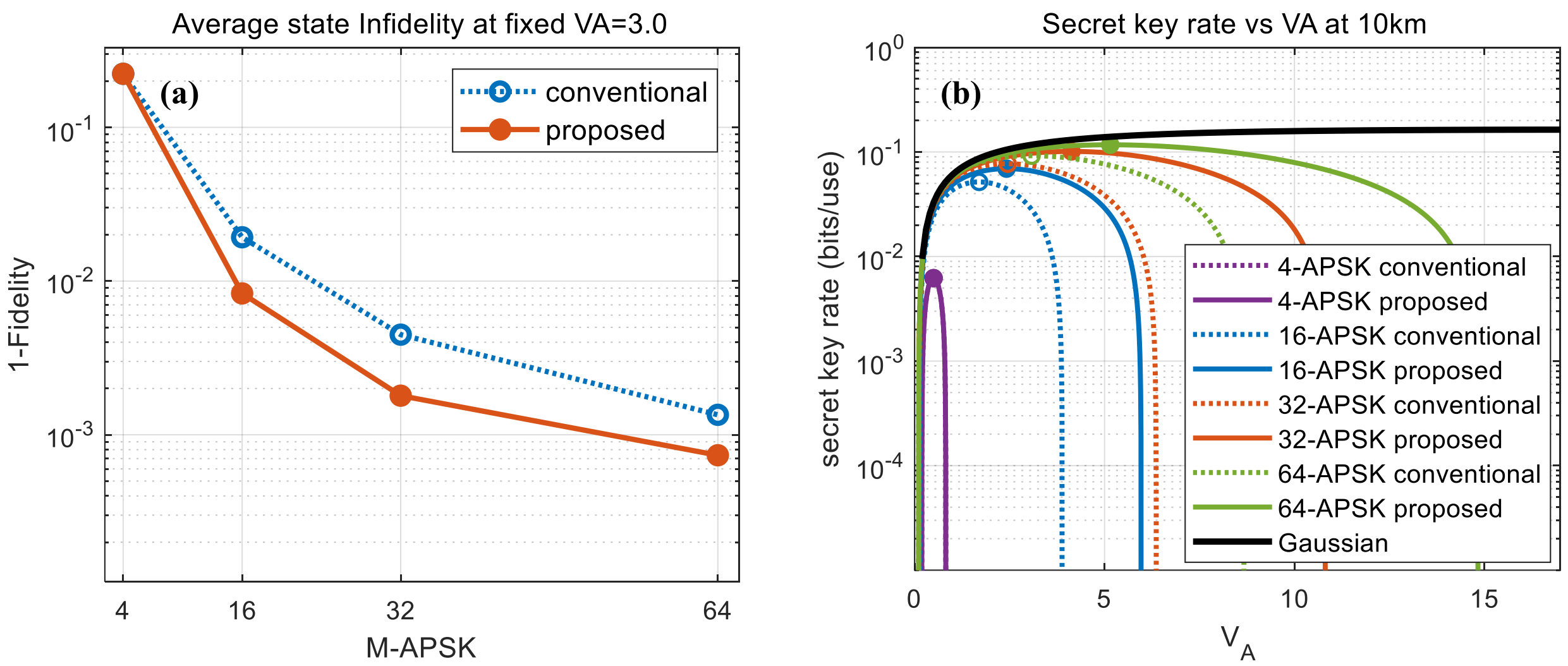

**Fig. 5** Multi-ring APSK analysis with increasing constellation size

(a) Infidelity comparison before and after optimization for different $M$.

(b) $V_A$-sweep Secret key rate curves for finding the optimal modulation variance at $L = 10$ km.

Fig. 5(a) compares the infidelity $1 - F(\tau_D, \tau_G)$ of the conventional binomial structure and the proposed $r, p$-optimized structure as a function of $M$. As $M$ increases, $1 - F$ decreases, which indicates that larger multi-ring APSK structures are already close to the Gaussian average state. Therefore, the remaining room for further optimization becomes smaller. The absolute reduction of $1-F$ after optimization is largest for 16-APSK, with a value of $1.09\times10^{-2}$. For 32-APSK and 64-APSK, the corresponding reductions are $2.7\times10^{-3}$ and $6.0\times10^{-4}$, respectively. This result shows that the proposed $r,p$-optimization is more effective for small size APSK structures, where the structural gap from Gaussian modulation is larger.

Fig. 5(b) shows the finite-size secret key rate obtained by sweeping $V_A$ at $L=10$ km. As $M$increases, the optimal $V_A$ also increases, and the range of $V_A$ with a positive secret key rate becomes wider. In addition, the proposed $r,p$-optimized structure shows a larger optimal $V_A$ and a wider positive secret key rate region than the conventional binomial structure.

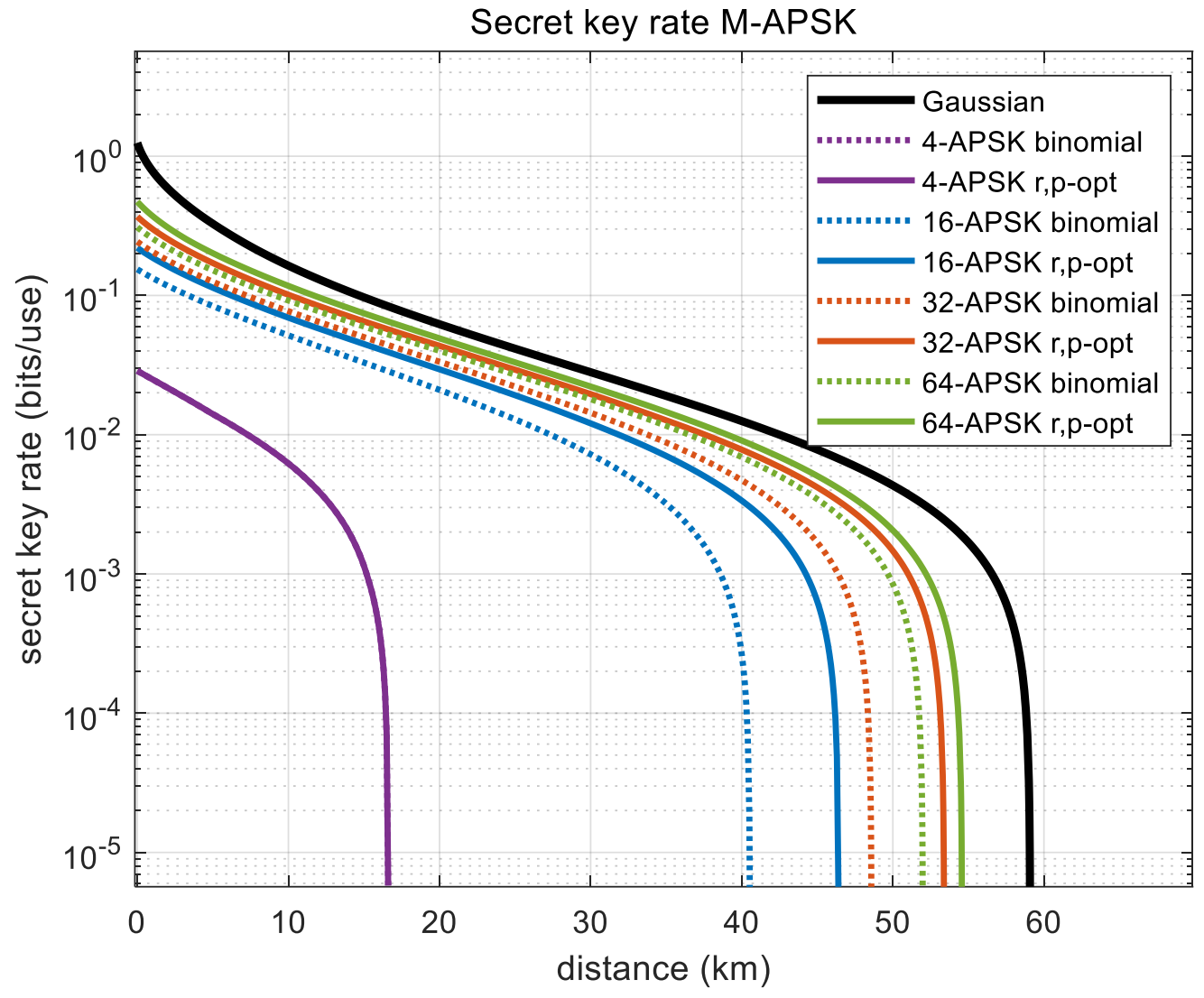

**Fig. 6** Finite-size secret key rate of multi-ring $M$-APSK as a function of transmission distance. The conventional binomial structure and the proposed $r,p$-optimized structure are compared.

**Table. 2** Maximum transmission distance comparison.

| Modulation | Conventional | Proposed | Distance gain | Improvement |
|---|---|---|---|---|
| 4-APSK (4) | 16.6 km | 16.6 km | 0.0 km | 0% |
| 16-APSK (4+12) | 40.5 km | 46.6 km | 6.1 km | 15.06% |
| 32-APSK (4+12+16) | 48.5 km | 53.8 km | 5.3 km | 10.93% |
| 64-APSK (4+12+16+32) | 51.9 km | 55.5 km | 3.6 km | 6.94% |

Fig. 6 and Table 2 show the improvement in the maximum transmission distance obtained by $r,p$-optimization for multi-ring APSK. The proposed structure achieves a longer transmission distance than the conventional binomial structure for all multi-ring APSK cases. However, the relative improvement decreases from 15.06% for 16-APSK to 10.93% for 32-APSK and 6.94% for 64-APSK. This is because, as $M$ increases, the conventional multi-ring APSK structure is already close to the Gaussian average state, and therefore the remaining room for improvement by additional ring radius and probability adjustment becomes smaller.

# 5 Conclusion

This paper proposed a multi-ring $M$-APSK constellation optimization method for discrete-modulated CV-QKD. The goal was to obtain an average state and finite-size secret key rate close to Gaussian modulation, while using a limited number of constellation points. The proposed approach consists of Gram matrix-based average-state calculation, ring radius and probability optimization, and fidelity-based structural analysis.

To compute the average state of multi-ring APSK, we used a Gram matrix-based method. Since the calculation of the correlation parameter $Z^*$ in Eq. (2) requires $\tau^{1/2}$ and $\tau^{-1/2}$, the spectral decomposition of the average state $\tau$ is needed. Instead of directly diagonalizing the truncated Fock-space matrix, the non-zero spectrum of $\tau$ was obtained from the $M \times M$ Gram matrix, as shown in Eqs. (10)-(16). This allows the eigen structure of the average state to be constructed more directly for multi-ring APSK constellations.

The ring radius and ring probability, which are fixed in the conventional multi-ring APSK structure, were then treated as optimization variables. As expressed in Eq. (20), for each transmission distance $L$, the modulation variance $V_A$, ring radius ratio $r$, and ring probability $p$ were jointly searched to maximize the finite-size secret key rate. For 16-APSK (4+12), the proposed $r, p$-optimization increased the maximum transmission distance by approximately 15% relative to the conventional binomial structure at the operational threshold $K_{\text{th}} = 1 \times 10^{-5}\ bits/use$.

To analyze the structural effect of the optimization, the fidelity in Eq. (25) was also used. The parameter $Z^*$ is directly related to the secret key rate, but it includes the effects of channel transmittance, excess noise, and the $w$ term. In contrast, fidelity directly compares the structural similarity between the discrete average state $\tau_D$ and the Gaussian average state $\tau_G$. As shown in Fig. 3, the optimized structures showed higher fidelity and larger $Z^*$, which is consistent with the improvement in the secret key rate. Therefore, the proposed $r, p$-optimization can be interpreted as a structural adjustment that makes $\tau_D$ closer to $\tau_G$.

The proposed method was further applied to 32-APSK and 64-APSK structures. As shown in Table 2 and Fig. 6, the proposed structure achieved a longer maximum transmission distance than the conventional binomial structure for all multi-ring APSK cases. However, the relative improvement decreased from approximately 15% for 16-APSK to 10.93% for 32-APSK and 6.94% for 64-APSK. This indicates that as $M$and the number of rings increases, the conventional multi-ring APSK structure is already close to the Gaussian average state, leaving less room for further improvement through radius and probability optimization.

Overall, this work shows that Gram matrix-based average-state calculation, ring radius and probability optimization, and fidelity-based structural analysis can be combined to improve and interpret the performance of multi-ring APSK modulation in DM-CV-QKD. Future work should consider the limitation of using a fixed reconciliation efficiency $\beta$, as well as more general discrete constellation optimization beyond the restricted APSK structure.